\documentclass[11pt,a4paper]{article}
\usepackage{jheppub}
\usepackage{graphicx}
\usepackage{amsmath}
\usepackage{multicol,bbm}

\pdfoutput=1

\newcommand{\be}{\begin{equation}}
\newcommand{\ee}{\end{equation}}
\newcommand{\bs}{\begin{split}}
\newcommand{\es}{\end{split}}
\newcommand{\bea}{\begin{eqnarray}}
\newcommand{\eea}{\end{eqnarray}}

\newcommand{\gsim}{\gtrsim}
\newcommand{\lsim}{\lesssim}

\newcommand{\beqa}{\begin{eqnarray}}
\newcommand{\eeqa}{\end{eqnarray}}

\newcommand{\vo}{\mathcal{V}}

\newcommand\fverb{\setbox\fverbbox=\hbox\bgroup\verb}
\newcommand\fverbdo{\egroup\medskip\noindent
			\fbox{\unhbox\fverbbox}\ }
\newcommand\fverbit{\egroup\item[\fbox{\unhbox\fverbbox}]}
\newbox\fverbbox

\begin{document}

\title{Towards Natural Inflation in String Theory}

\abstract{
We provide type IIB string embeddings of two axion variants of natural inflation. We use a combination of RR 2 form axions as the inflaton field and have its potential generated by non perturbative effects in the superpotential. Besides giving rise to inflation, the models developed take into account the stabilization of the compact space, both in the KKLT and large volume scenario regimes, an essential condition for any semi-realistic model of string inflation.
}

\preprint{DESY-14-118}

\author[a]{Ido Ben-Dayan,}
\author[a]{Francisco G. Pedro,} 
\author[a]{Alexander Westphal}

\affiliation[a]{Deutsches Elektronen-Synchrotron DESY, Theory Group, D-22603 Hamburg, Germany}
\emailAdd{ido.bendayan@desy.de}\emailAdd{francisco.pedro@desy.de}\emailAdd{alexander.westphal@desy.de}

\maketitle

\section{Introduction}

Recent observational progress has drastically transformed cosmology into a quantitative science \cite{Hinshaw:2012fq, Ade:2013zuv, Hou:2012xq, Sievers:2013wk}. From these measurements we derive increasingly strong evidence for cosmological inflation, a very early epoch of accelerated expansion lasting for  about 60 e-foldings of the scale factor.

Recently, the BICEP2 collaboration reported the first measurement of B-mode polarization of the CMB at large angular scales \cite{Ade:2014xna}. If this result stands after further corroboration and turns out to be primordial, then in the context of inflation it corresponds to a detection of primordial gravitational waves with a tensor-to-scalar ratio $r=0.16^{+0.06}_{-0.05}$. Further results quantifying the polarized foreground emissions from e.g. galactic dust along the lines of~\cite{Ade:2014gna,Flauger:2014qra,Mortonson:2014bja} will help settle this question in future.

The amount of e-folds of slow-roll inflation
\be
N_e=\int\limits_{\phi_e}^{\phi_{N_e}}\frac{d\phi}{\sqrt{2\epsilon}}
\ee
can be related to the tensor-to-scalar ratio, $r$, via the Lyth bound
\be
r=16\epsilon\sim 0.003 \left(\frac{50}{N_e}\right)^2 \left(\frac{\Delta\phi_{N_e}}{M_{\rm P}}\right)^2\quad .
\ee
If we take the B-mode detection for the time being as signalling a primordial inflationary tensor modes with $r\gtrsim 0.1$, then the Lyth bound \cite{Lyth:1996im,Boubekeur:2005zm} tells us that the field excursion during inflation was super-Planckian: $\Delta\phi_{N_e}\gg M_{\rm P}$ \footnote{The field excursion can be reduced to sub-Planckian, if we allow for a non-monotonic $\epsilon$, \cite{BenDayan:2009kv}. These types of models have received revived interest due to the BICEP2 result. We wish to note that there is no problem with having enough e-folds even with $\Delta \phi_{N_e} \ll 1$ and $n_s\simeq 0.96$ at the pivot scale per se. The actual limitation is that a smaller field excursion implies a faster change in $\epsilon$ which in turn implies a larger deviation from nearly scale invariance, which is harder to accommodate given the now $\simeq7$ e-folds measured by PLANCK \cite{BenDayan:2009kv,Hotchkiss:2011gz,Hebecker:2013zda}}.

Inflation is known to be sensitive to the high-scale effects of a possible UV completion of the low energy theory, however large and small field models are affected to different extents by this UV sensitivity. In small-field inflation $\Delta\phi_{N_e}\lesssim M_{\rm P}$ clearly requiring the tuning of dimension-6 operators to avoid ${\cal O}(1)$ contributions to the slow-roll parameter $\eta=V''/V$. To see this note that  a generic inflationary model will contain dimension-6 operators of the type $\delta V_6\sim V_0(\phi)\;\phi^2/M_{\rm P}^2$. In a small-field model the starting-point inflaton potential necessarily has the form
\be
V_0(\phi)=V_0\,\left(1+\sqrt{2\epsilon_0}\frac{\phi}{M_{\rm P}}+\frac{\eta_0}{2}\,\frac{\phi^2}{M_{\rm P}^2}+\ldots\right)\simeq V_0=const.
\ee
at small field values $\phi\ll M_{\rm P}$. %\ib{OMIT? Due to the Lyth bound the small-field range implies $\epsilon_0\ll\eta_0\ll 1$.} 
Hence, $\delta V_6\sim V_0\,\phi^2/M_{\rm P}^2$ corrects $\eta$ by an ${\cal O}(1)$ value, destroying slow-roll.

 Large-field inflation in contrast is UV sensitive to an infinite series of \emph{dangerously irrelevant} operators. Clearly, we must appeal to a protective symmetry to save the inflation direction in the scalar potential. This will almost by definition amount to an effective shift symmetry which is broken at leading order by the inflationary scalar potential 
itself. While this notion of a protective shift symmetry in large-field inflation is certainly natural in the bottom-up Wilsonian sense (corrections from self-interactions of inflaton fluctuations die out at large field values, and quantum Einstein gravity correction scale $\sim V / M_{\rm P}^4\; , \; V''/M_{\rm P}^2 \ll 1$), realizing such a shift symmetry and establishing control over its breaking clearly requires high-scale information mandating the embedding into a theory of quantum gravity.

At present there is not a unique and well understood theory of quantum gravity. However, string theory constitutes the most prominent candidate, with many non-trivial intricate results concerning its mathematical structure, the right low-energy field content to potentially accommodate our local universe, and a successful microscopic description of a large part of black hole physics. This provides a clear motivation to study inflation and in particular its  large-field varieties in string theory. The requirement of realizing a well-respected shift symmetry typically leads us to consider the many string theory axions from higher-dimensional $p$-form gauge fields, or their mirror-dual partners of complex structure moduli, as good inflaton candidates. However, many sectors of the theory display a periodicity under shift of the $p$-form axionic fields $a_{(p)}\to a_{(p)}+2\pi$ while the kinetic terms of these axions
\be
{\cal L}_{kin.}=f^2(\partial_\mu a_{(p)})^2
\ee
imply
\be
f\sim \frac{M_{\rm P}}{L^p}\ll M_{\rm P}
\ee
for 10d to 4d compactifications with volume ${\cal V}=L^6$ in string units. The periodicity range of the canonically normalized axion field
\be
\frac{\phi^{(p)}}{M_{\rm P}}=\frac{f}{M_{\rm P}} a_{(p)}\sim \frac{a_{(p)}}{L^p}\lesssim 1
\ee
is sub-Planckian in controllable compactifications requiring large volume and weak string coupling \cite{Banks:2003sx}. This discrete shift symmetry with sub-Planckian period is broken by the presence of quantized $p$-form fluxes or (dual) brane configuration, which unwrap the discrete shift symmetry into a system of multiple {\it non-periodic} branches: the full system with fluxes or branes shows {\it monodromy} in the potential energy of the axion on each branch, while periodicity is retained when summing over multiple branches \cite{Silverstein:2008sg, McAllister:2008hb,Kaloper:2008fb,Kaloper:2011jz,Palti:2014kza,Kaloper:2014zba,Marchesano:2014mla,Blumenhagen:2014gta,Hebecker:2014eua,Grimm:2014vva,Dine:2014hwa,McAllister:2014mpa}.

An alternative to monodromy proper arises in the presence of at least 2 axions~\cite{Kim:2004rp,Berg:2009tg,Choi:2014rja,Higaki:2014pja,Kappl:2014lra,Ben-Dayan:2014zsa,Tye:2014tja,Long:2014dta,Gao:2014uha,Li:2014lpa}. Non-perturbative effects provide cosine potentials with typical sub-Planckian periodicities $2\pi f_1$ and $2\pi f_2$
\be\label{eq:Vaxion}
V=\Lambda_1^4 \left[1-\cos\left(\frac{p_1}{f_1}\phi^{(p)}_1+\frac{p_2}{f_2}\phi^{(p)}_2\right)\right]+\Lambda_2^4 \left[1-\cos\left(\frac{q_1}{f_1}\phi^{(p)}_1+\frac{q_2}{f_2}\phi^{(p)}_2\right)\right]\quad .
\ee 
An alignment of the $p_i,q_i$ \cite{Kim:2004rp} or a hierarchy like e.g. $q_1\ll p_1,q_2$ with $p_2=0$ \cite{Ben-Dayan:2014zsa,Tye:2014tja} (a fully non-perturbative variant of \cite{Berg:2009tg}) drives the emergence of a mass hierarchy in the axionic sector, which is translated into the appearance of an effective super-Planckian axion decay constant $f_{eff.}\gg M_{\rm P}\gtrsim f_1, f_2$. The arising effective single-field inflaton potential realizes the original idea of natural inflation~\cite{Freese:1990rb}.

A crucial step for any construction of large-field inflation in string theory consists of showing the compatibility of the shift symmetry and field-range extension mechanism with the process of moduli stabilization. The moduli potential tends to back react on the inflationary vacuum energy. This generates corrections to the inflaton potentials, which energetically lead generically to flattening of a naive potential shape as inferred from the pure large-field mechanism itself \cite{Dong:2010in}. Generically, the inflationary vacuum energy may very well participate in moduli stabilization which often can enhance the stability of the compactification while showing the flattening effect  \cite{Dong:2010in, McAllister:2014mpa}. If we restrict ourselves to the supersymmetric setup provided by Calabi-Yau flux compactification, stabilization of the volume moduli often requires non-perturbative effects. For this reason, CY compactifications often lead to problems with having the inflationary vacuum energy participating in moduli stabilization, requiring a (moderate) separation of scales between the moduli potential and the inflaton sector. This, however, is an artifact of our restricting to CY compactifications in the first place.

In this paper we are discussing several methods of embedding an aligned or hierarchical axion potential of the type~\eqref{eq:Vaxion} into type IIB string theory compactified on CY manifolds with 3-form flux fixing the complex structure moduli and the axio-dilaton \cite{Giddings:2001yu}. Due to the arguments discussed above, and those expressed in section \ref{sec:axionsinstring},  we will restrict ourselves to the use 2-form R-R sector axions $C_2$ which provide a rather well-protected shift symmetry in the context of type IIB on CY orientifolds with O7 planes and D3/D7 branes \cite{Grimm:2004uq, Grimm:2007xm, McAllister:2008hb}. 

The paper is organized as follows. In section $2$, we discuss natural inflation models in the two axions case. We will demonstrate that \cite{Kim:2004rp,Ben-Dayan:2014zsa,Tye:2014tja,Berg:2009tg} actually stem from the same origin. We embed such models in a supergravity (SUGRA) framework  in section $3$.  In section $4$ we discuss their string theoretic derivation. We will study both the KKLT mechanism \cite{Kachru:2003aw} and the large-volume scenario \cite{Balasubramanian:2005zx} for combining K\"ahler moduli stabilization with non-perturbative effects from gaugino condensation on $D5$-branes or $ED3-ED1$ instanton from Euclidean D3-branes providing the axion potential for the 2 R-R sector $C_2$ axions. Finally we conclude in section $5$.

\section{Natural inflation from two axions}\label{sec:FT}
\paragraph{A common origin\\}
In this section we discuss the two mechanisms that generate effective super-Planckian decay constants from fundamental sub-Planckian ones, following the works \cite{Kim:2004rp,Ben-Dayan:2014zsa,Tye:2014tja}. We show that actually both models (as well as \cite{Berg:2009tg}) come from the same origin and only correspond to different deformations of the underlying potential, or different breaking of the same shift symmetry.

Consider a two axion potential of the form 
\be
V=\Lambda_1^4\left(1-\cos\left(\frac{p_1}{\tilde{f}_1}\phi_1+\frac{p_2}{\tilde{f}_2}\phi_2\right)\right)+\Lambda_2^4\left(1-\cos\left(\frac{q_1}{\tilde{f}_1}\phi_1+\frac{q_2}{\tilde{f}_2}\phi_2\right)\right),
\label{eq:VKNP}
\ee
where the axions $\phi_1$ and $\phi_2$ have canonical kinetic terms and all decay constants are sub-Planckian: $\frac{\tilde{f}_1}{p_1},\frac{\tilde{f}_2}{p_2},\frac{\tilde{f}_1}{q_1},\frac{\tilde{f}_2}{q_2}< M_P$.

We start our discussion by imposing an alignment condition on the above scalar potential which ensures the presence of flat direction by arranging for $V=V(\phi_1+\phi_2)$. 
Then the determinant of the second derivative matrix vanishes everywhere, signalling the possibility of a flat direction
$\det V_{ij}=0$ \footnote{In Mathematics $\det V_{ij}=0$ is called the Monge-Amp\`ere equation. Starting from such an equation is much more general than our discussion here. The Monge-Amp\`ere equation provides an excellent starting point for model building and model classification. We intend to return to this in future work.}.
Hence all that is left is to find a region where the gradient $V_i$ is small, and we have the flat region desired for inflation, provided that the other eigenvalues of $V_{ij}$ are positive. This happens naturally at the origin of field space in \eqref{eq:VKNP}.
Slow-roll inflation requires deforming the flat direction slightly. We can achieve this in two non-equivalent directions away from the flat limit -- either we relax the alignment condition in the spirit of Kim-Nilles-Peloso, or we introduce a subdominant scalar potential providing the slowly varying deformation away from flatness as in Dante's inferno and Hierarchical Axion inflation. We will now discuss both possibilities in more detail.

\paragraph{KNP\\}

We start by discussing the Kim-Nilles-Peloso (KNP) mechanism. As was mentioned above, in the perfect alignment limit $\frac{p_2}{p_1}=\frac{q_2}{q_1}$ the mass matrix is singular. With inflation model building in mind one deforms this condition to allow for small misalignment:
\be
\frac{p_2}{p_1}\equiv r \qquad\text{and}\qquad \frac{q_2}{q_1}\equiv r (1+\delta)\qquad\text{with}\qquad \delta\ll 1.
\label{eq:misalignment}
\ee
This has the effect of lifting the flat direction in a way suitable for slow-roll inflation provided $\delta$ is sufficiently small. One must note that for sufficiently small $\delta$, the alignment mechanism holds for generic values of the various parameters in Eq. (\ref{eq:VKNP}), as can be seen from the fact that the determinant of the mass matrix
\be
\det M^2\equiv m_1^2 m_2^2= \Lambda_1^4\Lambda_2^4 \frac{ p_1^2 q_1^2 r^2 }{\tilde{f}_1^2 \tilde{f}_2^2}\delta^2 
\ee
becomes singular in the limit of perfect alignment ($\delta=0$), signalling the presence of a flat direction.

To leading order in $\delta$, the mass eigenvalues are
\be
m_1^2= \frac{p_1^2 q_1^2 r^2}{\tilde{f}_2^2+\tilde{f}_1^2 r^2} \frac{\Lambda_1^4 \Lambda_2^4}{p_1^2 \Lambda_1^4+ q_1^2 \Lambda_2^4} \delta^2 \qquad\text{and}\qquad m_2^2=\frac{\tilde{f}_2^2 +\tilde{f}_1^2 r^2}{\tilde{f}_1^2 \tilde{f}_2^2} \left(p_1^2 \Lambda_1^4 + q_1^2 \Lambda_2^4\right)
\ee
which keeping in mind that the axions' masses are of the form $\Lambda^4/f^2$ implies that the large effective decay constant, corresponding to the eigenvalue $m_1^2$, scales as 
\be
f_{eff}^2= \frac{\tilde{f}_2^2+\tilde{f}_1^2 r^2} {q_1^2 r^2 \delta^2}
\label{eq:feffKNP}
\ee 

And so, by considering two almost aligned axions with originally sub-Planckian decay constants, one generates an effective super-Planckian decay constant as is required for natural inflation for the price of tuning an alignment of the original axion decay constants.
This extension of the field range without requiring a super-Planckian fundamental domain constitutes the main advantage of this model, when compared with the original single--cosine realization of natural inflation \cite{Freese:1990rb}.

\paragraph{Hierarchical Axions\\}
The Hierarchical Axions (HA) model (and its close cousin Dante's Inferno) corresponds to setting $p_2=0$ in \eqref{eq:VKNP}. 
By definition there is no alignment whatsoever and the deformation of the potential that will allow for inflation corresponds to  introducing a hierarchy in the decay constants by having  $p_1\ll q_1$. The mass matrix at the global minimum is such that 
\be
\det M^2=\Lambda_1^4\Lambda_2^4 \frac{ p_1^2 q_2^2 }{\tilde{f}_1^2 \tilde{f}_2^2},
\ee
so in the limit $p_1\rightarrow 0$ an exact shift symmetry is recovered and for $q_1\gg p_1$, the symmetry is broken such that inflation ends at a stable minimum at the origin, provided that
\be
\label{hierarch}
\frac{\Lambda_2}{\Lambda_1}>\sqrt\frac{p_1}{q_1}.
\ee
%Considering the case of $\Lambda_1\ll \Lambda_2$, the axions masses 
The effective axion masses in the limit $q_1\gg p_1$ are
 %In the limit of small $f_{r_2}\ll f_{r_1}, f_{\theta_2}$ the eigenvalues of $M^2$ are
\be
m_1^2=\frac{\Lambda _1^4}{\tilde {f}_2^2}\left(\frac{p_1 q_2}{q_1}\right)^2,\qquad m_2^2=\frac{\Lambda _1^4p_1^2}{\tilde{f}_1^2}+\Lambda _2^4\left(\frac{q_1^2}{\tilde{f}_1^2}+\frac{q_2^2}{\tilde{f}_2^2}\right) 
\ee
and so we see that the mass spectrum is hierarchical, with  $m_1\ll m_2$. Integrating out the heavy mode results in a single effective axion potential with an effective decay constant:
\be
\label{eq:feffHA}
f_{eff}=\tilde{f}_2 \frac{q_1}{q_2 p_1}
\ee
The model generates super-Planckian effective decay constants similarly to KNP, but replaces the tuned alignment with a hierarchy between the decay constants. It further keeps the entire inflationary analysis in a sub-Planckian domain avoiding the functional fine-tuning necessary in large field models.
The advantages of the model as laid out in \cite{Ben-Dayan:2014zsa} are threefold:
i) utilizing only non-perturbative effects, ii) the smallest
number of axions, iii) and the least amount of tuning of
the input parameters. These advantages make the model specifically tractable from the string theoretic point of view. %Specifically, in other models such as DI or axion monodromy, there is the concern about the magnitude of backreaction of the brane on the background geometry, and whether that invalidates the model completely or at least changes its predictions considerably. The HA model does not share that concern since it is made up of strictly non-perturbative effects. \\

Both the KNP alignment and the Hierarchical Axions mechanism produce an effectively single-field inflaton potential of the form $V(\phi)=\Lambda_{eff.}^4\,\left[1-\cos(\phi/f_{eff.})\right]$ with an effective enhanced field range $f_{eff.}\gtrsim M_{\rm P}$. Hence, its observational predictions for the spectral index $n_s$ of curvature perturbation, and the tensor-to-scalar ratio $r$ agree with those of natural inflation itself %\cite{Savage:2006tr, 
\cite{Kappl:2014lra}. In particular, for $f_{eff.}\gtrsim 10 M_{\rm P}$ the scalar potential in the 60 e-fold range approaches that of $m^2\phi^2$-inflation with $n_s=1-2/N_e\simeq 0.967$ and $r=8/N_e\simeq 0.13$, while for $f_{eff.}\lesssim 5 M_{\rm P}$ the model becomes of small-field type with $n_s$ growing more red and leaving the Planck 95\% region, while the tensor-to-scalar ratio drops to (currently) unobservable levels ($r\to 0$ for $f_{eff.}\to 0$).

%We present the embedding of these models in a SUGRA setting in section 3 and in string theory setup which includes moduli stabilization in section 4.

\section{Supergravity embeddings}

In this section we provide explicit ways to build the field theory models  \cite{Kim:2004rp,Ben-Dayan:2014zsa,Tye:2014tja} described above  into supergravity. The goal of this approach is to provide a stepping stone to the realisation of these inflationary models in string compactifications of type IIB string theory which we present in section \ref{sec:string}.

The minimal model requires  two chiral multiplets $X_m\equiv  b_m+i\ c_m$ whose dynamics are determined by the canonical K\"ahler potential 
\be
K=\frac{1}{4} \left(X_1+\overline{X}_1\right)^2+\frac{1}{4} \left(X_2+\overline{X}_{2}\right)^2
\label{eq:Ksugra2}
\ee
and the non-perturbatively generated superpotential 
\be
W=W_0+A e^{-p_1 X_1-p_2 X_2}+B e^{-q_1 X_1-q_2 X_2}.
\label{eq:Wsugra2}
\ee

In the case $p_2\neq0$ one has the KNP alignment mechanism giving rise to a super-Planckian axion, while if one chooses from the start $p_2=0$ inflation will proceed through hierarchies in the parameters $p_1, q_1 \text{ and } q_2$. The structure of the scalar potential will be similar in both cases and so we analyse them simultaneously whenever possible.

The scalar potential can be made to have a hierarchy between the terms stabilising the real parts of $X_1$ and $X_2$ and those generating the inflationary potential, thereby decoupling the heavy fields from the inflationary dynamics. 

The F term potential
\be
V=e^{K}\left(D_I W D^I\bar{W} -3 |W|^2\right), \qquad\text{where}\qquad D_I W=\partial_IW+W\partial_I K
\label{eq:VF}
\ee

 can be written as 
\be
V=V_0+V_1+V_2
\ee
where
\be
V_0= e^{b_1^2+b_2^2} W_0^2 \left(-3+2 b_1^2+2 b_2^2\right),
\ee
\be
V_1= 2 e^{b_1^2+b_2^2}\ W_0 \left \{ A\ F_2[p_1, p_2]\ \cos\left[p_1 c_1+p_2 c_2\right] +B\ F_2[q_1, q_2]\ \cos\left[q_1 c _1+q_2 c _2\right] \right \}
\label{eq:V2SUGRA}
 \ee
and
\be
\begin{split}
V_2=&A^2 F_3\left[p_1,p_2\right]+B^2 F_3\left[q_1,q_2\right]+2 A B e^{- \left(p_1+q_1\right)b_1-\left(p_2+q_2\right) b_2+b_1^2+b_2^2} \cos\left[(p_1-q_1 )c _1+\left(p_2-q_2\right) c _2\right]\\
&\times \left(-3+2 \left(p_1-b_1\right) \left(q_1-b_1\right)+2 \left(p_2-b_2\right) \left(q_2-b_2\right)\right)
\end{split}
\ee 
For the sake of short formulae we have defined the $b_1$ and $b_2$ dependent quantities 
\be
\begin{split}
F_2[m,n]&\equiv e^{-m b_1-n b_2} \left(-3+2 b_1 \left(-m+b_1\right)+2 b_2 \left(-n+b_2\right)\right)\ ,\\
F_3[m,n]&\equiv e^{-2 m b_1+b_1^2-2 n b_2+b_2^2} \left(-3+2 \left(m-b_1\right){}^2+2 \left(n-b_2\right){}^2\right)\ .
\end{split}
\ee

Both in the KNP and the hierarchical regimes we focus on regions of parameter space where $|W_0|\gg |A e^{-p_1 X_1}|,|A e^{-p_1 X_1-p_2 X_2}|,|B e^{-q_1 X_2 -q_2 X_2}|$ so that one can stabilise $b_1, b_2$ at  high scale before analysing the inflationary potential. This hierarchy in the superpotential descends into the scalar potential: $V_0\gg V_1 \gg V_2$.   In these simple supergravity models the need for this hierarchy derives from the desire to analytically minimise the potential in a controlled way and it is not a fundamental requirement of these models since we expect the alignment/hierarchical mechanisms to produce a super-Planckian direction even in the absence of this hierarchy in $W$. 
This situation will change once we consider stringy embeddings of this idea, as we will see in section \ref{sec:string}, since this tuning will be related to parametric decoupling of the moduli vacuum. 

We note in passing, that more generally parametric decoupling is an overly conservative criterion which may me relaxed in concrete string embeddings. The moduli potential may  back react appreciably during the inflaton evolution, which due to energetic reasons generically leads to a flattening of the inflaton potential expected from the parametrically decoupled limit~\cite{Dong:2010in}. In particular, if perturbative high-scale mechanisms serve to fix all the moduli, inflation may even participate and help with moduli stabilization allowing for significant yet controllable flattening effects~\cite{McAllister:2014mpa}. The non-perturbative mechanisms of volume stabilization we use here are more sensitive to backreaction effects. This limits the amount of controllable flattening achievable, and motivates us to restrict ourselves to the limit of parametric decoupling for the sake of explicitness.

Note, that the combined minimum of the three cosine terms above and the moduli potential usually is an AdS vacuum. As discussed in many of the dS vacuum constructions in string theory in recent years, we need to add an uplifting contribution from e.g. an anti D3-brane \cite{Kachru:2003aw}, D-terms with field-dependent FI terms \cite{Burgess:2003ic, Cicoli:2012vw} or dilaton dependent non perturbative effects \cite{Cicoli:2012fh}
\be
\delta V_{\rm uplift}=\frac{C}{{\cal V}^p}\;\;,\;\;{\rm with}\;\; p = {\cal O}(1) >0
\ee
to the scalar potential to lift the AdS minimum to a near Minkowski state $\langle V_F+\delta V_{\rm uplift}\rangle\simeq 0$. Provided, we already arranged for sufficient hierarchy between the moduli masses and the axion mass scales arising from the three cosine terms in $V_F$ in the prior AdS vacuum, this will survive an uplifting term of the above type. Hence, we will from now on tacitly assume the presence of such an uplifting term in our setups, which justifies the form $V(\phi)\sim 1-\cos(\phi/f)$ for the three cosine terms arising from the moduli potential.

The structure of the F-term potential, Eq. (\ref{eq:VF}) implies that a three cosine potential is inevitable in supergravity, modifying Eq. (\ref{eq:VKNP}) to
\be\begin{split}
V&= \Lambda_1^4\left(1-\cos\left(\frac{p_1}{\tilde{f}_1}\phi_1+\frac{p_2}{\tilde{f}_2}\phi_2\right)\right)+\Lambda_2^4\left(1-\cos\left(\frac{q_1}{\tilde{f}_1}\phi_1+\frac{q_2}{\tilde{f}_2}\phi_2\right)\right)\\
&+\Lambda_3^4\left(1-\cos\left(\frac{p_1}{\tilde{f}_1}\phi_1+\frac{p_2}{\tilde{f}_2}\phi_2-\frac{q_1}{\tilde{f}_1}\phi_1-\frac{q_2}{\tilde{f}_2}\phi_2)\right)\right).
\end{split}
\label{eq:3CosV}
\ee
Even though this will alter the expressions for the mass eigenvalues, which will receive $\Lambda_3$ dependent contributions, the existence of a mass hierarchy remains and the large effective decay constants are still given by Eqs. (\ref{eq:feffKNP}) and \eqref{eq:feffHA} for the KNP and HA cases respectively.

\begin{figure}[h!]
	\centering
	\begin{minipage}[b]{0.48\linewidth}
	\centering
	\includegraphics[width=\textwidth]{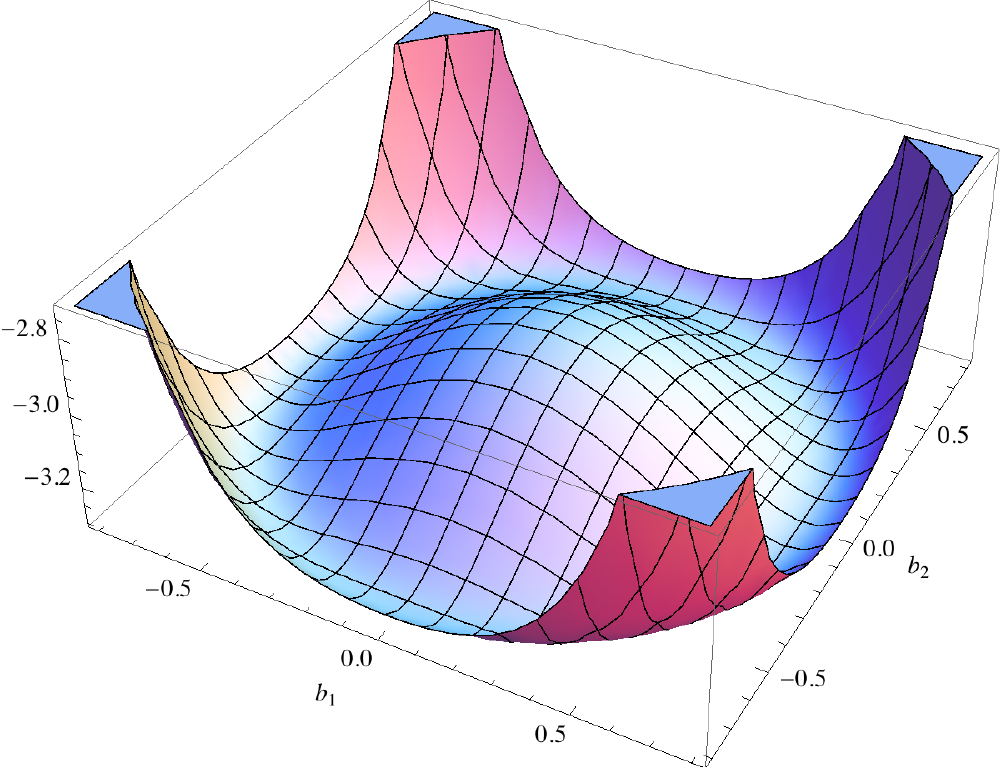}
    \end{minipage}
	\hspace{0.1cm}
	\begin{minipage}[b]{0.48	\linewidth}
	\centering
	\includegraphics[width=\textwidth]{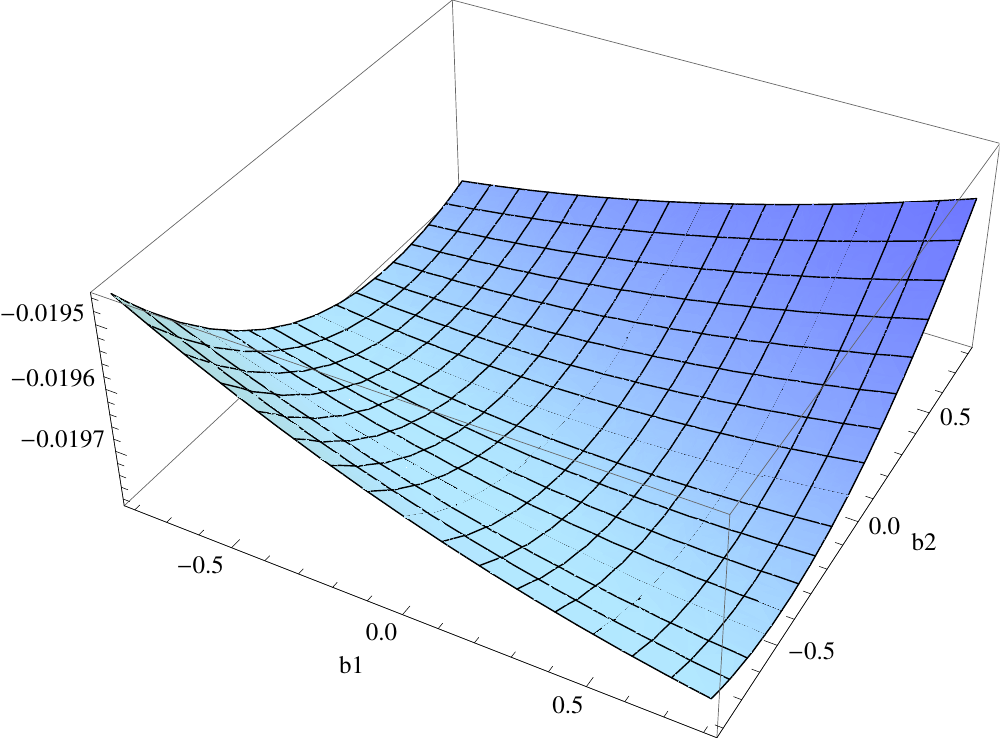}
	\end{minipage}
	\caption{Scalar potential in the $(b_1,b_2)$ plane for $\{A,B, W_0, r, \delta, b_1, a_1\}=\{  0.002, 0.001, 1,  2, 0.1, 0.1, 0.05\}$. Right: $V_0$, left: $V_1$. Note the hierarchy between the two contributions to $V$.}
	\label{fig:sugraII}
\end{figure}

 Noting that  $V_0$ depends only on the combination $y\equiv b_1^2+b_2^2$ we see that the tree level action stabilises $y$ at
 \be
\langle y \rangle=1/2,
 \ee
and so $b_1, b_2$ must lie in a circle of radius $1/ \sqrt{2}$ centred at the origin of the $(b_1, b_2)$ plane. This is illustrated in figure \ref{fig:sugraII}. At this level there is still one flat direction left in the  $(b_1, b_2)$ plane as well as two in the $(c_1,c_2)$. Since $V_0$ is independent from $p_2$ this result applies to both the KNP and the hierarchical axion mechanism. The flat directions will be lifted by the next-to-leading order contribution to the potential,  $V_1$, which we analyze separately in the two regimes.

\subsection{KNP alignment mechanism}

We start by studying the vacua of $V_1$ in KNP in the limit of perfect alignment, $\frac{p_2}{p_1}=\frac{q_2}{q_1}\equiv r$, where we know there will be one unfixed direction in the $c$-plane. We will then follow \cite{Kim:2004rp} and allow for a slight misalignment which will lift the remaining flat direction in a way suitable for inflation. This procedure is more subtle here than in the  field theory case since the leading contribution to the potential, $V_0$ leaves the angular direction in the $(b_1,b_2)$ plane unfixed. One therefore has to ensure that by allowing for a slight misalignment in the decay constants in order to realise inflation, one is not simultaneously destabilising the angular direction in the $(b_1,b_2)$. 

In the alignment limit we have 
\be
V_1=-4 W_0 e^{\frac{1}{2}-\frac{z \left(p_1+b_1\right)}{p_1}}\left( A e^{\frac{z q_1}{p_1}} (1+z) \cos\left[p_1 \left(c _1+r c _2\right)\right] +B e^z \left(1+z \frac{q_1}{p_1}\right)  \cos\left[q_1 \left(c _1+r c _2\right)\right] \right )
\ee
where for algebraic simplicity we have set $p_1=q_1$ and defined $z\equiv p_1 b_1+p_2 b_2$. This potential will simultaneously stabilise $z$ and the axionic combination $c_1+r c_2$. The minimum is located at $\langle z \rangle=0$ and $\langle c_1+r c_2 \rangle=0$. This implies that the vacuum in the $b$-plane lies at the intersection of the circle $b_1^2 + b_2^2=1/2$ and the straight line passing through the origin $p_1 b_1+p_2 b_2=0$. By construction, in this limit, the same linear combination of $c_1, c_2$ appears in the potential and so the orthogonal combination is exactly flat. 

The introduction of a slight misalignment $\delta$ in the axionic decay constants, 
\be
\frac{p_2}{p_1}\equiv r\qquad\text{and}\qquad\frac{q_2}{q_1}\equiv r (1+\delta),
\ee
which will generate the inflationary potential, will also perturb the b-vacuum. This can in extreme cases lead to the destruction of the b-minimum or in more mild ones lead to a shift in the minimum's position. Since for inflationary purposes $\delta\ll1$, %one can show that 
 the b-vacuum survives the introduction of a misalignment, with its position shifting by a small factor proportional to the smallness of the misalignment of the decay constants $\delta$:
\be
\langle z \rangle=-\delta \frac{B r p_1 q_1^2 }{\sqrt{2} \sqrt{1+r^2} \left(A p_1^2+B q_1^2\right)}.
\ee
This fixes the remaining flat direction in the $b-$plane in the misaligned regime in a similar way to what was described before. The only difference is that the straight line intersecting the circle in the $b$-plane no longer passes through the origin. The effect of the misalignment in the stabilisation of the real part of the fields $X_1$ and $X_2$ is therefore negligible.

The potential for the axions then admits the following expansion
\be
V_1=-4 W_0  \sqrt{e} \left\{A \cos\left[p_1 \left(c _1+r c _2\right)\right] +  B \cos\left[q_1 \left(c _1+r (1+\delta ) c _2\right)\right]\right\}
\ee
which coincides with the field theory model of Eq. (\ref{eq:VKNP}), with the identifications $\Lambda_1^4= 4 W_0 A \sqrt{e}$ and $\Lambda_2^4= 4 W_0 B \sqrt{e}$.

\subsection{Hierarchical axions mechanism}

Just like in the KNP case, for the hierarchical axion inflation model, the potential is generated by Eq. (\ref{eq:V2SUGRA}), but now with $p_2=0$. 
Anticipating that inflationary model building will require $q_1\gg q_2, p_1$ we can expand $V_1$ as
\be
V_1= -4 W_0 \left(A \sqrt{e} \cos[p_1 c_1]+ B e^{1/2+z}(z-1)\cos[q_1c_1+ q_2 c_2]\right),
\ee
where $z\equiv -q_1 b_1 -q_2 b_2$.

 $V_1$ is simultaneously responsible for stabilising the angular direction in the b-plane (or equivalently stabilising $z$) and giving rise to inflation. To leading order in a $q_1$ expansion, we find that $\langle z\rangle=0$, implying that the b-vacuum is the same as in the aligned KNP regime.  In this case the potential then simplifies to the desired form
\be
V_1=-4 W_0 \left(A \sqrt{e} \cos[p_1 c_1]+ B \sqrt{e} \cos[q_1c_1+ q_2 c_2]\right).
\ee

One then concludes that the model of Eqs.  (\ref{eq:Ksugra2}), (\ref{eq:Wsugra2}) does indeed provide  a supergravity description of the two axion versions of natural inflation of \cite{Kim:2004rp} and \cite{Ben-Dayan:2014zsa,Tye:2014tja}. Generically this did not have to be the case since the structure of supergravity requires a model that initially involves four fields: 2 axions and 2 saxions/moduli, whose masses are closely linked. What we have shown is that  the mechanism that allows for the generation of a potential suitable for inflation simultaneously guarantees that there is a mass hierarchy between the lightest axion and all the saxions/moduli. When trying to embed these models in string compactifications we will therefore be looking for the generic structures of (\ref{eq:Ksugra2}), (\ref{eq:Wsugra2}), the adequate mass hierarchies and tuneable parameters.

\section{Natural inflation in string compactifications}
\label{sec:string}

\subsection{Axions in string compactifications}\label{sec:axionsinstring}
We confine our discussion to GKP-type compactifications of type IIB string theory on O3/O7 orientifolds of warped Calabi-Yau 3-folds with 3-form NS-NS and R-R flux. We assume a choice of 3-form fluxes such that they stabilize the complex structure moduli and the type IIB axio-dilaton supersymmetrically at a high mass scale while generating an effectively constant superpotential $W_0$~\cite{Giddings:2001yu}.

For the stabilization of the K\"ahler moduli we consider non-perturbative stabilization a la KKLT~\cite{Kachru:2003aw} involving gaugino condensation on D7-brane or D5-brane stacks or Euclidean D3-brane (ED3) instantons, or the Large Volume Scenario (LVS) involving a combination of the leading ${\cal O}(\alpha'^3)$-correction to the CY 3-volume and an ED3-brane instanton or D7-brane stack.

At the ${\cal N}=2$ level prior to imposing the O7 projection, the K\"ahler moduli sector consists of $h^{1,1}$ 2-cycle moduli $t^j=v^j+i b^j$. Here the $b^j=\int_{\Sigma_{2_j}}B_2$ denotes the NS-NS 2-form axions $b^j$ arising from the NS 2-form $B$-field on the $h^{1,1}$ 2-cycles $\Sigma_{2_j}$, while the $v_j$ denote the 2-cycle geometric volumes in string units. The tree-level Calabi-Yau volume ${\cal V}$ is then given by
\be
{\cal V} = \frac16\,k_{ijk} v^i v^j v^k\quad.
\ee
Imposing the O7 projections projects the K\"abler moduli sector into an O7-even and odd subspaces with respective dimensions $h^{1,1}_+$ and $h_-^{1,1}=h^{1,1}-h^{1,1}_+$. Moreover, this forces a rearrangement of the real scalars into $a=1\dots h^{1,1}_+$ 4-cycle moduli
\be
T_a=\frac12\,k_{abc}v^b v^c+i\int_{\Sigma_4^a}C_4 +\frac{1}{2(S+\bar S)}k_{a\beta\gamma}G^\beta (G^\gamma+\bar G^\gamma)
\ee
and $\alpha=1\ldots h^{1,1}_-$ 2-form axion multiplets
\be
G^\alpha=\bar S b^\alpha+i c^\alpha
\ee
where the type IIB axio-dilaton is $S=e^{-\phi}+iC_0$ and we have O7-odd 2-form R-R and NS-NS axions
\be
c^\alpha=\int_{\Sigma_{2_\alpha}} C_2 \quad {\rm and}\quad b^\alpha=\int_{\Sigma_{2_\alpha}} B_2\quad.
\ee
For the case of a single K\"ahler modulus $h^{1,1}_+=1$ and $h^{1,1}_-\geq 1$ odd 2-form axion moduli we can invert the relation between $T$ and $t$ and acquire $t(T,G^\alpha)$. This gives a 4d ${\cal N}=1$ K\"ahler potential of the K\"ahler moduli (see e.g. \cite{Lust:2006zg})
\be\label{Kk}
K=-3\ln\left[T+\bar T+\frac{1}{2(S+\bar S)}k_{1\beta\gamma}(G+\bar G)^\beta (G^\gamma+\bar G^\gamma)\right]\quad.
\ee
Guided by this example, we conjecture that for CY manifolds with a volume of swiss-cheese form with $h^{1,1}_+>1$ K\"ahler moduli we may find cases where the K\"ahler potential takes the form (see e.g. \cite{Gao:2013rra})
\bea
K=-2\ln{\cal V}\;,\;{\rm with}\quad {\cal V}&=&c_L\left(T_L+\bar T_L +\frac{1}{2(S+\bar S)}k_{L\beta\gamma}(G+\bar G)^\beta (G^\gamma+\bar G^\gamma)\right)^{3/2}\\
&&-\hspace{-1ex}\sum\limits_{i=2\ldots h^{1,1}_+}\hspace{-1ex}c_i\left(T_i+\bar T_i +\frac{1}{2(S+\bar S)}k_{i\beta\gamma}(G+\bar G)^\beta (G^\gamma+\bar G^\gamma)\right)^{3/2}\;\;.\nonumber
\eea
The swiss-cheese type 4-cycle intersection numbers then are positive $c_L,c_i >0$. Due to the mixing between the $G^\alpha$ and $\tau$ in $K$ the results for the full $(T_a, G^\alpha, \tau)$-sector K\"ahler metric and its inverse are bit lengthy. The full expressions are given in eq.s (C.11) and (C.12) of Appendix C in~\cite{Grimm:2004uq}.

We can now see from the form of e.g. Eq.~\eqref{Kk} that the projection dictated by the O7-action leads to a breaking of the shift symmetry which the $B_2$-axions enjoyed at the 4D ${\cal N}=2$ level. If moduli stabilization proceeds via terms in the superpotential, then $W$ must be a holomorphic function of the chiral ${\cal N}=1$ superfields. Hence, superpotential stabilization of the moduli generically stabilizes the $T_a$ and $G^\alpha$ separately. The manifest dependence of $K$ on the $B_2$-axions being $\sim {\rm Im}\,G^\alpha$ hence breaks the shift symmetry of the $B_2$-axions. This fact renders the NS-NS 2-form axions unsuitable for large-field inflation in these constructions.

If volume moduli stabilization had proceeded directly via stabilizing some of the geometric 4-cycle volumes e.g. by corrections to $K$, this could have restored the $B_2$ shift symmetry for the $G^\alpha$ involved. However, we will not pursue this opportunity here, and in our constructions the $G^\alpha$ will always appear in $W$ non-perturbatively.

This leaves the $C_4$ and $C_2$ R-R axions as potential inflaton candidates. Non-perturbative stabilization a la KKLT of the $T_a$ K\"ahler moduli then implies that the $C_4$-axions $\sim {\rm Im}\,T_a$ acquire the same mass scales as their moduli partners ${\rm Re}\, T_a$. In the Large Volume Scenario we require only a fraction of the $T_a$ to appear in $W$ via non-perturbative effects. For those which do not appear in $W$ but get fixed by K\"ahler corrections, this allows to split the mass scale of their $C_4$-axion partners from their moduli mass scale~\cite{Cicoli:2014sva}. In such cases, we may use $C_4$-axions as inflaton candidates.

Generically, however, we find the $C_2$-axions are least coupled to the process of volume stabilization regardless whether this proceeds non-perturbatively in $W$ or perturbatively in $K$. Hence, we focus on 2-axion models of large-field inflation driven by the R-R $C_2$ axions of two $G^\alpha$ multiplets.

\subsection{Embedding into string compactifications}

In this section we try to build  explicit models of the KNP and HA mechanisms using the orientifold odd axions of string compactifications.  This analysis is a little more involved than the supergravity case since not only do we have to generate a suitable potential for the axions, but we must also stabilise the geometry of the compact space. To achieve this we combine the KKLT or LVS setups with superpotential terms originating from gaugino condensation on D5 branes~\cite{Grimm:2007xm,Grimm:2007hs}:

\be
W=W_0+A e^{-a T} + P e^{- p_1 G_1 - p_2 G_2}+Q e^{- q_1 G_1 - q_2 G_2},
\label{eq:W}
\ee
where $T$ can be chosen as needed between the even moduli in each setup.

\subsubsection{Inflating in KKLT}

In the simplest case of KKLT moduli stabilisation, we consider a single even modulus model, such that the volume of the compact space is given by $\mathcal{V}= t^3$, where $t$ is the 2-cycle volume. Allowing for non-zero intersection between the orientifold even and the odd sectors, one can write the K\"ahler potential in terms of the K\"ahler coordinates as 
\be
K=-3\log \left[ T+\bar{T} +  \frac{k_1}{2(S+\bar{S})}(G_1+\bar{G_1})^2 + \frac{k_2}{2 (S+\bar{S})}(G_2+\bar{G_2})^2\right]-\log\left[S+\bar{S}\right].
\label{eq:K}
\ee
The kinetic part of the scalar field Lagrangian then reads
\be
\mathcal{L}_{kin}= K_{I\bar{J}}\partial_\mu \Psi_I\partial^\mu \Psi_{\bar{J}}=\kappa_{ i j}\partial_\mu \psi_i\partial^\mu \psi_{i},
\ee
where $\Psi$ denotes the chiral superfield basis, $\Psi =\{ T, G_1, G_2,S\}$,  and $\psi$ the real scalar field basis of moduli space $\psi=\{ \tau, \rho, b_1,c_1,b_2, c_2, s, C_0\}$. The kinetic matrix for the real degrees of freedom, ignoring the dilaton for the moment,  admits the following expansion
\be
\kappa_{ij}=\left(
\begin{array}{cccccc}
 \frac{3}{4 \tau^2} & 0 & 0 & 0 & 0 & 0 \\
 0 &  \frac{3}{4 \tau^2} & 0 & 0 & 0 & 0 \\
 0 & 0 & -\frac{3 k_1}{4 g_s \tau} & 0 & 0 & 0 \\
 0 & 0 & 0 & -\frac{3 k_2}{4 g_s \tau} & 0 & 0 \\
 0 & 0 & 0 & 0 & -\frac{3 g_s k_1}{4 \tau} & 0 \\
 0 & 0 & 0 & 0 & 0 & -\frac{3 g_s k_2}{4 \tau}
\end{array}\right),
\label{eq:metricKKLT}
\ee
where we keep only the leading terms in $\vo$ in each diagonal entry and neglected off-diagonal terms\footnote{ The full expression of the kinetic matrix up to order $1/\vo^2$ is somewhat complicated and not very illuminating. By presenting only the leading diagonal terms one can understand the behaviour of the eigenvalues of $\kappa_{ij}$ in terms of the intersection numbers $k_1$ and $k_2$ and get a simple estimate of their order of magnitude.}. We then see that requiring positivity of $\kappa_{ij}$ leads us to consider compactifications with $k_1, k_2<0$. If these conditions are not met, the $G$ multiplets become ghosts.

The scalar potential resulting from Eqs. (\ref{eq:W}) and (\ref{eq:K}) can be written as  $V=V_0+V_1+V_2$, which after minimising the 4-form axion at $\rho=\frac{-\pi +a (b_1 c_1 k_1+ b_2 c_2 k_2)}{a}$ become
\be
\begin{split}
V_0=&\frac{9 W_0^2 \left(b_1^2 k_1+b_2^2 k_2\right){}^2}{64 \tau ^5 g_s}\\
& + A W_0 e^{-a \tau +\frac{a b_1^2 k_1}{2 g_s}+\frac{a b_2^2 k_2}{2 g_s}} \left(-\frac{a g_s}{4 \tau ^2}-\frac{9 \left(b_1^2 k_1+b_2^2 k_2\right){}^2}{32 \tau ^5 g_s}+\frac{3 a \left(b_1^2 k_1+b_2^2 k_2\right){}^2}{16 \tau ^4 g_s}\right)\\
& + A^2 e^{-2 a \tau +\frac{a b_1^2 k_1}{g_s}+\frac{a b_2^2 k_2}{g_s}} \left( \frac{a^2 g_s}{12 \tau }+\frac{9 \left(b_1^2 k_1+b_2^2 k_2\right){}^2}{64 \tau ^5 g_s}-\frac{3 a \left(b_1^2 k_1+b_2^2 k_2\right){}^2}{16 \tau ^4 g_s}+\frac{a^2 \left(b_1^2 k_1+b_2^2 k_2\right){}^2}{16 \tau ^3 g_s}\right.\\ &\left.-\frac{a \left(-3 g_s+a b_1^2 k_1+a b_2^2 k_2\right)}{12 \tau ^2}\right),
\end{split}
\label{eq:V0kklt}
\ee

\be
\begin{split}
V_1=&e^{-\frac{b_1 p_1}{g_s}-\frac{b_2 p_2}{g_s}} P \cos\left[c_1 p_1+c_2 p_2\right] \ \times\\
&\left[-\frac{a A e^{-a \tau +\frac{a \left(b_1^2 k_1+b_2^2 k_2\right)}{2 g_s}} g_s}{4 \tau ^2}-\frac{9 A e^{-a \tau +\frac{a \left(b_1^2 k_1+b_2^2 k_2\right)}{2 g_s}} \left(b_1^2 k_1+b_2^2 k_2\right){}^2}{32 \tau ^5 g_s}+\frac{9 W_0 \left(b_1^2 k_1+b_2^2 k_2\right){}^2}{32 \tau ^5 g_s}\right.\\
&\left.+\frac{3 a A e^{-a \tau +\frac{a \left(b_1^2 k_1+b_2^2 k_2\right)}{2 g_s}} \left(b_1^2 k_1+b_2^2 k_2\right){}^2}{16 \tau ^4 g_s}-\frac{a A e^{-a \tau +\frac{a \left(b_1^2 k_1+b_2^2 k_2\right)}{2 g_s}} \left(b_1 p_1+b_2 p_2\right)}{6 \tau ^2}\right.\\
&\left.-\frac{3 A e^{-a \tau +\frac{a \left(b_1^2 k_1+b_2^2 k_2\right)}{2 g_s}} \left(b_1^2 k_1+b_2^2 k_2\right) \left(b_1 p_1+b_2 p_2\right)}{8 \tau ^4 g_s}+\frac{3 W_0\left(b_1^2 k_1+b_2^2 k_2\right) \left(b_1 p_1+b_2 p_2\right)}{8 \tau ^4 g_s}\right.\\
&\left.+\frac{a A e^{-a \tau +\frac{a \left(b_1^2 k_1+b_2^2 k_2\right)}{2 g_s}} \left(b_1^2 k_1+b_2^2 k_2\right) \left(b_1 p_1+b_2 p_2\right)}{4 \tau ^3 g_s}\right] +  \Big[P\rightarrow Q, p_1\rightarrow q_1, p_2\rightarrow q_2 \Big]
\end{split}
\label{eq:V1kklt}
\ee
and
\be\begin{split}
V_2\approx&-P^2 \frac{\left(k_2 p_1^2+k_1 p_2^2\right)}{12 \tau ^2 k_1 k_2}e^{\frac{2 \left(b_1 p_1+b_2 p_2\right)}{g_s}}-Q^2 \frac{\left(k_2 q_1^2+k_1 q_2^2\right)}{12 \tau ^2 k_1 k_2}e^{\frac{2 \left(b_1 q_1+b_2 q_2\right)}{g_s}}\\
&-P Q \frac{  \left(k_2 p_1 q_1+k_1 p_2 q_2\right)}{6 \tau ^2 k_1 k_2}e^{\frac{b_1 \left(p_1+q_1\right)+b_2 \left(p_2+q_2\right)}{g_s}} \cos\left[c_1 \left(p_1-q_1\right)+c_2 \left(p_2-q_2\right)\right].
\end{split}
\label{eq:V2kklt}
\ee
Note that while the Eqs. (\ref{eq:V0kklt}) and (\ref{eq:V1kklt}) are exact, for the sake of short formulae in Eq.  (\ref{eq:V1kklt}) we have displayed only the first non-vanishing terms in a $1/\tau$ expansion.

A successful string inflation model must not only give rise to inflation but also be able to keep the non-inflationary moduli fixed. In the current context this requires a separation between the physical mass scales of the  $\tau$, $\rho$ $b_1$ and $b_2$ fields and the $c_1$ and $c_2$ axions. This scale separation is also a prerequisite for the classical stability of the vacuum: after uplifting the KKLT vacuum of $V_0$ is separated from decompactification by a barrier with a height $V_{barrier}\sim |V_{KKLT}|\sim \frac{W_0^2}{\vo^2}$. Since any inflationary energy density constitutes an extra form of uplifting, one must have $V_{inf}<V_{barrier}\sim  \frac{W_0^2}{\vo^2}$. The twin requirements of scale separation and vacuum stability therefore impose the following hierarchy on the scalar potential
\be
V_{0}\gg V_1+V_2,\qquad\text{implying}\qquad P, Q \ll \frac{W_0}{\sqrt{\tau}}.
\ee
Provided this is met one can  minimise $V_0$ and $V_1+V_2$ separately, which constitutes a considerable simplification in the search for the F-term potential's vacuum.

The leading contribution to the potential, $V_0$, depends only on the volume modulus $\tau$ and on the quadratic combination $b_1^2 k_1+b_2^2 k_2$, and is essentially the generalisation of the  KKLT potential for compactifications with orientifold odd axions intersecting the volume modulus. Extremising $V_0$ we find that the KKLT minimum is approximately located at 
\be
e^{-a \tau +\frac{a \left(b_1^2 k_1+b_2^2 k_2\right)}{2 g_s}} \approx \frac{3 W_0}{2 a A \tau}\left(1+\frac{-3+4 a\frac{b_1^2 k1 + b_2^2 k_2}{2 g_s}}{2 a \tau}\right).
\label{eq:MinV0}
\ee
As in the supergravity versions of these two axion models, the real partners of the axionic fields are K\"ahler stabilised at leading order in a circle of fixed radius, determined by the solution to $\frac{\partial V_0}{\partial b_i}=0$\footnote{Due to the structure of the potential we have not been able to find an analytic expression for the vev of $k_1 b_1^2+k_2 b_2^2$ that gave a good agreement with the numerical results while still being compact enough to be spelled out here. We therefore proceed with the analysis numerically, keeping in mind that in the cases of interest one finds $\frac{a \left(b_1^2 k_1+b_2^2 k_2\right)}{2 g_s} \sim \mathcal{O}(-\text{a few})$, in accordance with the requirement $k_1,k_2<0$, derived from the abscence of ghosts.}, with the angular direction unfixed.

The $V_1+V_2$ component of the scalar potential depends both on $\tau$ and $b_1^2 k_1+b_2^2 k_2$ as well as on various linear combinations of $b_1$ and $b_2$ and on the $c$ axions, in both the KNP and hierarchical axions scenarios, and so $V_1+V_2$ will lift  the remaining flat direction in the $b$-plane. Unfortunately, the structure of the potential complicates the minimisation process as soon as one moves away from the (tachyonic) origin of the b-plane. This renders our efforts to find analytic expressions for the location of the $(b_1,b_2)$ vacuum futile and forces us to resort to numeric methods. In any case, the qualitative picture is identical to that of the supregravity models described in the previous section, and at the end of the process one ends up with all the moduli, volume included, stabilised in a consistent way.

\paragraph{KNP alignment mechanism\\}

At the KKLT minimum, the dominant contribution to $V_1$ is
\be
V_1\approx -P W_0 \frac{3 g_s-2 \left(b_1 p_1+b_2 p_2\right) }{8 \tau ^3}e^{\frac{b_1 p_1+b_2 p_2}{g_s}}  \cos\left[c_1 p_1+c_2 p_2\right]+ \Big[P\rightarrow Q, p_1\rightarrow q_1, p_2\rightarrow q_2 \Big], \label{eq:V1kkltknp}
\ee
which together with the cosine term from $V_2$, Eq. (\ref{eq:V2kklt}), constitute the inflationary potential $V_{inf}$.

Defining the misallignemt parameter  $\delta$ in terms of the superpotential parameters as 
\be
\frac{p_2}{p_1}\equiv r\qquad\text{and}\qquad\frac{q_2}{q_1}\equiv r (1+\delta).
\ee
one can map this stringy model onto the field theory analysis of section \ref{sec:FT}. The potential for the canonically normalised $c$-axions
\be
\phi_{1}=\tilde{f}_{1}\ c_{1}\qquad\text{and}\qquad\phi_{2}=\tilde{f}_{2}\ c_{2}
\label{eq:phi}
\ee
 can then be written, upon uplifting, as
\be\begin{split}
V_{inf}& = \Lambda_1^4 \left(1-\cos\left[\frac{\phi_1}{\tilde{f}_1/p_1}+ \frac{\phi_2}{\tilde{f}_2/( p_1 r)}\right]\right)+ \Lambda_2^4 \left(1-\cos\left[\frac{\phi_1}{\tilde{f}_1/q_1}+ \frac{\phi_2}{\tilde{f}_2/(q_1 r(1+\delta))}\right]\right)\\
&+\Lambda_3^4\left(-1 \cos\left[\frac{\phi_1}{\tilde{f}_1/\left(p_1-q_1\right)} + \frac{\phi_2}{\tilde{f}_2/\left(r p_1-(1+\delta ) r q_1\right)} \right]\right),
\end{split}
\ee
where $\Lambda_{1},\Lambda_{3} \text{ and }\Lambda_{3}$ can be read off Eqs. (\ref{eq:V2kklt}) and (\ref{eq:V1kkltknp}) and therefore scale as
\be
\Lambda_1^4\sim \mathcal{O}\left(\frac{W_0 P }{\tau^3}\right),\qquad\Lambda_2^4\sim \mathcal{O}\left(\frac{W_0 Q }{\tau^3}\right),\qquad \Lambda_3^4\sim \mathcal{O}\left(\frac{ P Q }{\tau^2}\right).
\ee
%\be
%\begin{split}
%\Lambda_{1,}^4=& -\frac{e^{\frac{\left(r b_1+b_2\right) p_2}{g_s}} W_0\left(3 g_s-2 \left(r b_1+b_2\right) p_2\right) X_{1,2}}{8 \tau ^3}+ \mathcal{O}(\delta)\\
%\sim&\  \mathcal{O}\left(\frac{W_0 \ X_{1,2}}{\tau^3}\right)
%\end{split}
%\ee
From Eq. (\ref{eq:metricKKLT}) one also defines
\be
\tilde{f}_{1,2}= \sqrt{\frac{-3 k_{1,2} g_s}{2 \tau}}
\ee
and so one finds that the large effective decay constant is then written in terms of the compactification parameters as 
\be
%f_{eff}= \frac{\tilde{f_2}}{r\ p_2\ \delta}\ .
f_{eff}= \frac{\sqrt{\tilde{f_2}^2+r^2 \tilde{f_1}^2}}{r\ q_1\ \delta}\ .
\ee
If there are no large asymmetries neither in the intersection numbers, $k_1\sim k_2$, nor in the superpotential exponents, $r\sim 1$, one finds $f_{eff} \sim \frac{\tilde{f}_1}{q_1\delta}$. Working in a region of parameter space where $k_1\sim k_2\sim -\mathcal{O}(1)$, $g_s\ll1$ and $\tau\sim \mathcal{O}(10)$\footnote{ As is often the case in models of  string inflation, the compactification volume is closely tied with the scale of inflation, which is in turn subject to observational bounds on the amplitude of the curvature perturbations. We therefore expect models with significantly larger $\tau$ to be at least extremely contrived, if not inconsistent with observations. } one tends to find $\tilde{f}_{1,2}\lsim \mathcal{O}(0.1)$. Furthermore noting that since the $G$ moduli potential is generated by gaugino condensation on D5 branes implies $q_1=\frac{2 \pi}{N}$, one sees that for $f_{eff}\gsim 5 M_P$ one must have the misalignment parameter tuned at the level of 
\be
\delta \lsim\mathcal{O}\left(\frac{N}{300}\right).
\ee

There are several possible avenues to arrange for tuning the alignment parameter into a range of $\delta = 0.01\ldots 0.1$. One possibility is to employ CY manifolds which contain non-trivial linear combinations of the symplectic-basis 2-cycles (Poincar\`e dual to so-called 'partially ample' 4-cycle divisors). The gauge kinetic function of the 5-branes wrapping these linear-combination 2-cycles then depends on a linear combination $k_1G_1+k_2G_2$ with $k_1,k_2$ numbers which are related to the intersection numbers determining the non-trivial linear-combination 2-cycle. By sampling over such 2-cycles and CYs containing these type of linear-combination 2-cycles, we should be able to find examples where $k_2=k_1(1+{\cal O}(\delta))$ providing the necessary level of alignment. 

Alternatively, we may use D7-brane gaugino condensates (or ED3-ED1 instantons) instead of D5-branes, and then turn on magnetic $F_2$ flux on them~\cite{Long:2014dta}. The D7-gauge kinetic function will then be $f_{a,D7}=T_a+f^iG_i+\ldots$ with $f^i$ being the quanta of $F_2$-flux possible on the 4-cycle the D7 stack wraps, and $G_i$ those $G$-axion fields which appear in the D7-Chern-Simons action once the $f^i$ are turned on. In this case, tuning the magnetic flux quanta may allow for reaching the necessary alignment $f^2=f^1(1+{\cal O}(\delta))$. We note further, that the $\exp(-aT)$ prefactors present in the non-perturbative terms from the D7-brane stacks or ED3-ED1s contribute part of the tuning of the effective instanton prefactors $P,Q$ similar to what we find in section~\ref{sec:LVS} for Large Volume Scenario. In that case we expect values of the residual effective $P,Q\sim{\cal O}(0.01\ldots0.1)$.

Finally, in the tables \ref{tab:KNP_inkklt} and \ref{tab:KNP_out} we present two numerical examples of the model described above.

\begin{table}[h]
\begin{center}
\begin{tabular}{c||c|c|c|c|c|c|c|c|c|c|c|c}
& $W_0$ & $A$ & $a$ & $P$ & $Q$ & $p_1$&  $q_1$ & $r$ & $\delta $ & $k_1$ & $k_2$ & $g_s$\\
\hline
\hline
KNP$_1$ &$0.11$&$5$&$0.1$&$10^{-4}$&$10^{-4}$&$\pi/2$&$\pi/2$&$1$&$0.02$&$-3$&$-3$&$0.5$\\
\hline
KNP$_2$ &$0.099$&$1$&$0.1$&$10^{-4}$&$10^{-3}$&$\pi/5$&$\pi/5$&$1.5$&$0.1$&$-7$&$-7$&$0.3$\\
\end{tabular}
\end{center}
\caption{Input parameters for the string embedding of the KNP mechanism.}
\label{tab:KNP_inkklt}
\end{table}

\begin{table}[h]
\begin{center}
\begin{tabular}{c||c|c|c||c|c|c|c}
& $\tau$ & $b_1$ & $b_2$ & $\Lambda_1^4\times 10^{12}$&$\Lambda_2^4\times 10^{12}$ &$\Lambda_3^4\times 10^{12}$& $f_{eff}$\\
\hline
\hline
KNP$_1$ &$51$&$0.59$&$-0.59$&$11$&$11$&$0.98$&$9.4$\\
\hline
KNP$_2$ &$34$&$0.29$&$-0.18$&$18$&$180$&$2.8$&$5.8$\\
\end{tabular}
\end{center}
\caption{Compactification features and inflationary parameters. All dimensionful quantities expressed in Planckian units. We note that while we are presenting here only one minimum  for $(b_1,b_2)$, there are in general there are two degenerate minima.}
\label{tab:KNP_out}
\end{table}

\paragraph{Hierarchical axions mechanism\\}

The analysis of the inflationary potential of Eq. (\ref{eq:V1}) proceeds in a similar manner to the closely related KNP mechanism. By making the discrete choice $p_2=0$ and working in the limit $q_1\gg p_1, q_2$, we find
\be\begin{split}
V_1+V_2&= -P W_0\frac{3 g_s-2 b_1 p_1}{8 \tau ^3}e^{\frac{b_1 p_1}{g_s}}  \cos\left[c_1 p_1\right] -Q W_0\frac{ 3 g_s-2 \left(b_1 q_1+b_2 q_2\right)}{8 \tau ^3}e^{\frac{b_1 q_1+b_2 q_2}{g_s}} \cos\left[c_1 q_1+c_2 q_2\right]\\
&-P Q \frac{p_1 q_1}{6 \tau ^2 k_1} e^{\frac{b_1 p_1}{g_s}+\frac{b_1 q_1}{g_s}+\frac{b_2 q_2}{g_s}}  \cos\left[c_1 p_1-c_1 q_1-c_2 q_2\right] - P^2\frac{ p_1^2}{12 \tau ^2 k_1}e^{\frac{2 b_1 p_1}{g_s}}\\
&-Q^2\frac{ k_2 q_1^2+k_1 q_2^2}{12 \tau ^2 k_1 k_2}e^{\frac{2 b_1 q_1}{g_s}+\frac{2 b_2 q_2}{g_s}} 
\end{split}\ee

As before, $V_1+V_2$ is simultaneously responsible for the stabilisation of the linear combination $b_1 p_1 + b_2 p_2$ \footnote{We do not present an explicit analytic formula for the location of this minimum as it involves solving a non-algebraic equation.} and for the inflationary dynamics. After integrating out the $b$ fields and uplifting it reduces to
\be
V_{inf}=\Lambda_1^4 \left(1-\cos[ p_1 c_1 ]\right)+ \Lambda_2^4\left(1- \cos[ q_1 c_1+q_2 c_2]\right)+ \Lambda_3^4 \left(1-\cos[(p_1-q_1)c_1+ q_2 c_2]\right),
\label{eq:V1Simp}
\ee
where the leading scaling of the amplitudes is given by
\be
\Lambda_1^4\sim\mathcal{O}\left( \frac{P W_0}{\tau^3} \right),\qquad\Lambda_2^4\sim\mathcal{O}\left(\frac{Q W_0}{\tau^3} \right),\qquad\Lambda_3^4\sim\mathcal{O}\left( \frac{P Q}{\tau^2}\right).
\ee

%If the compactification is such that not only $W_0\gg X_1, X_2$ but also $W_0/\tau \gg X_1,X_2$ we find the hierarchy $\Lambda_1\sim\Lambda_2\gg\Lambda_3$, which allows us to neglect the last term of Eq. (\ref{eq:V1Simp}), reducing it to Eq. (hierarchicalInflationEq in sect 2) up to an overall  constant uplifting term. If the second condition does not hold, we'll have in general $\Lambda_1\sim\Lambda_2\sim\Lambda_3$, in which case one can still reduce the problem to the simple form of () by noting that to leading order $p_1-q_1\sim p_1$.

Noting that to leading order $p_1-q_1\sim -q_1$ the potential can be reduced to the simple form of (\ref{eq:VKNP}) where the amplitude of the mixed cosine is now $\Lambda_2^4+\Lambda_3^4$.  Writing the potential in terms of the canonically normalised fields, defined by Eq. (\ref{eq:phi}), one finds
\be
V_{inf}\approx \Lambda_1^4\left(1- \cos\left[ \frac{p_1}{\tilde{f_1}} \phi_1\right]\right)+ \left(\Lambda_2^4+\Lambda_3^4\right) \left(1-\cos\left[ \frac{q_1}{\tilde{f_1}} \phi_1+ \frac{q_2}{\tilde{f_2}} \phi_2 \right]\right),
\ee
implying that the large effective decay constant is
\be
f_{eff}=\tilde{f_2}\ \frac{q_1}{p_1 q_2}.
\ee

In tables \ref{tab:HA_inkklt} and \ref{tab:HA_out} we present a couple of numerical examples.

\begin{table}[h]
\begin{center}
\begin{tabular}{c||c|c|c|c|c|c|c|c|c|c|c}
& $W_0$ & $A$ & $a$ & $X_1$ & $X_2$ & $p_1$&  $q_1$ & $q_2$ & $k_1$ & $k_2$ & $g_s$\\
\hline
HA$_1$ & $0.022$ & $1$ & $0.1$& $0.0001$& $0.0001$& $\pi/10$& $\pi$& $\pi/10$& $-7$& $-7$& $0.4$\\
\hline
HA$_2$ & $0.0099$& $1$& $0.2$& $0.0001$& $0.0001$& $\pi/7$& $2\pi$& $\pi/7$& $-2$& $-2$&$0.4$\\
\end{tabular}
\end{center}
\caption{Input parameters for the string embedding of the hierarchical axions' mechanism.}
\label{tab:HA_inkklt}
\end{table}

\begin{table}[h]
\begin{center}
\begin{tabular}{c||c|c|c||c|c|c|c}
& $\tau$ & $b_1$ & $b_2$ & $\Lambda_1^4 \times 10^{12}$&$\Lambda_2^4\times 10^{12}$& $\Lambda_3^4\times 10^{12}$ & $f_{eff}$\\
\hline
\hline
HA$_1$ & $50$& $0.01$& $-0.49$& $1.7$& $1.6$& $0.06$&$9.1$\\
\hline
HA$_2$ & $30$& $-0.013$& $-0.71$& $4 $& $2.7$& $0.9$& $6.2$\\
\end{tabular}
\end{center}
\caption{Compactification features and inflationary parameters. All dimensionful quantities expressed in Planckian units. We note that while we are presenting here only one minimum  for $(b_1,b_2)$, there are in general there are two degenerate minima.}
\label{tab:HA_out}
\end{table}

\subsection{Inflating in LVS}\label{sec:LVS}

Having shown that effective trans-Planckian decay constants can coexist with stabilised moduli in the context of a single modulus KKLT compactification we now set out to do the same within the large volume scenario (LVS) of type IIB flux compactifications. 

Our starting point is a compactification which beyond the two orientifold odd $G$-fields giving rise to inflation also includes a pair of orientifold even $T$ moduli, whose triple intersection numbers lead to a Swiss cheese geometry. This setup implies the following K\"ahler potential
\be
K=-2\log \left[ \Sigma_1^{3/2}- \Sigma_2^{3/2}+\xi\ (S+\bar{S})^{3/2}\right]-\log\left[S+\bar{S}\right],
\label{eq:KLVS}
\ee
where
\be\begin{split}
&\Sigma_1\equiv T_1+\bar{T_1} +  \frac{k_{11}}{2 (S+\bar{S})}(G_1+\bar{G_1})^2 + \frac{k_{12}}{2 (S+\bar{S})}(G_2+\bar{G_2})^2,\\
&\Sigma_2\equiv T_2+\bar{T_2} +  \frac{k_{21}}{2 (S+\bar{S})}(G_1+\bar{G_1})^2 + \frac{k_{22}}{2 (S+\bar{S})}(G_2+\bar{G_2})^2.
\end{split}\ee
Given that we want to stabilise the moduli \'{a} l\'{a} LVS we have also included in $K$ the $\mathcal{O}(\alpha'^3)$ correction originating from the fourth order curvature correction to the ten dimensional action.

Before we analyse the F-term potential that follows from Eqs. (\ref{eq:W}) and (\ref{eq:KLVS}), we compute the kinetic matrix for the real degrees of freedom $\psi=\{\tau_1,\rho_1,\tau_2,\rho_2,b_1, b_2, c_1, c_2\}$ finding:
\be
\kappa_{ij}=\left(
\begin{array}{cccc|cccc}
 \frac{3}{4 \tau _1^2} & 0 & 0 & 0 & 0 & 0 & 0 & 0 \\
 0 & \frac{3}{4 \tau _1^2} & 0 & 0 & -\frac{3 c_1 k_{11}}{2 \tau _1^2} & -\frac{3 c_2 k_{12}}{2 \tau _1^2} & 0 & 0 \\
 0 & 0 & \frac{3}{8 \tau _1^{3/2} \sqrt{\tau _2}} & 0 & 0 & 0 & 0 & 0 \\
 0 & 0 & 0 & \frac{3}{8 \tau _1^{3/2} \sqrt{\tau _2}} & -\frac{3 c_1 k_{21}}{4 \tau _1^{3/2} \sqrt{\tau _2}} & -\frac{3 c_2 k_{22}}{4 \tau _1^{3/2} \sqrt{\tau _2}} & 0 & 0 \\\hline
 0 & -\frac{3 c_1 k_{11}}{2 \tau _1^2} & 0 & -\frac{3 c_1 k_{21}}{4 \tau _1^{3/2} \sqrt{\tau _2}} &    &    &    &    \\
 0 & -\frac{3 c_2 k_{12}}{2 \tau _1^2} & 0 & -\frac{3 c_2 k_{22}}{4 \tau _1^{3/2} \sqrt{\tau _2}} &    &   \tilde{\kappa}_{mn}  &    &    \\
 0 & 0 & 0 & 0 &    &   &    &    \\
 0 & 0 & 0 & 0 &    &    &    &   
\end{array}
\right)\ee
with
\be\begin{split}
&\tilde{\kappa}_{mn} =\\
&\left(
\begin{array}{cccc}
 -\frac{3 k_{11}}{4 g_s \tau _1}+\frac{3 c_1^2 k_{21}^2}{8 \tau _1^{3/2} \sqrt{\tau _2}}+\frac{3 k_{21} \sqrt{\tau _2}}{4 g_s \tau _1^{3/2}} & \frac{3 c_1 c_2 k_{21} k_{22}}{4 \tau _1^{3/2} \sqrt{\tau _2}} & 0 & 0 \\
 \frac{3 c_1 c_2 k_{21} k_{22}}{4 \tau _1^{3/2} \sqrt{\tau _2}} & -\frac{3 k_{12}}{4 g_s \tau _1}+\frac{3 c_2^2 k_{22}^2}{8 \tau _1^{3/2} \sqrt{\tau _2}}+\frac{3 k_{22} \sqrt{\tau _2}}{4 g_s \tau _1^{3/2}} & 0 & 0 \\
 0 & 0 & -\frac{3 g_s k_{11}}{4 \tau _1}+\frac{3 g_s k_{21} \sqrt{\tau _2}}{4 \tau _1^{3/2}} & 0 \\
 0 & 0 & 0 & -\frac{3 g_s k_{12}}{4 \tau _1}+\frac{3 g_s k_{22} \sqrt{\tau _2}}{4 \tau _1^{3/2}}
\end{array}
\right)
\end{split}
\label{eq:OddMetric}
\ee
where we have set, in anticipation of what is to come, $b_1=b_2=0$. Noting that the eigenvalues of $\kappa_{ij}$ are mostly determined by the diagonal entries, we see from Eq. (\ref{eq:OddMetric}) that for the orientifold odd moduli to have positive kinetic terms one must arrange 
\be
k_{11} \sqrt{\tau _1}-k_{21} \sqrt{\tau _2}<0\qquad\text{and}\qquad k_{12} \sqrt{\tau _1}-k_{22} \sqrt{\tau _2}<0.
\label{eq:ghosts}
\ee

The scalar potential that generalises LVS for $k_{+--}\neq 0$ takes the form
\be\begin{split}
V_{LVS}=&\frac{9 \left(b_1^2 k_{11}+b_2^2 k_{12}\right){}^2 W_0^2}{64 g_s \tau _1^5}+\frac{3 \xi  W_0^2}{32 \sqrt{g_s} \tau _1^{9/2}}-\frac{a^2 A^2 \left(b_1^2 k_{12} k_{21}^2+b_2^2 k_{11} k_{22}^2\right)}{12 k_{11} k_{12} \tau _1^2}e^{a \left(\frac{b_1^2 k_{21}+b_2^2 k_{22}}{g_s}-2  \tau _2\right)} \\
&+\frac{a^2 A^2 g_s \sqrt{\tau _2}}{6 \tau _1^{3/2}}e^{a \left(\frac{b_1^2 k_{21}+b_2^2 k_{22}}{g_s}-2\tau _2\right)} -\frac{g_s a A  W_0 \tau _2}{4 \tau _1^3}e^{a \left(\frac{b_1^2 k_{21}+b_2^2 k_{22}}{2 g_s}- \tau _2\right)}.\label{eq:VLVS}
\end{split}\ee

Solving $\partial V_{LVS}/\partial b_1=0$, $\partial V_{LVS}/\partial b_2=0$ one finds that the origin of the b-plane $(b_1, b_2)=(0,0)$ is an extremum of $V_{LVS}$. At this point in moduli space, the even moduli vacuum is determined in the usual way to lie at
\be
\tau _1^{3/2} =\frac{3 W_0 \sqrt{\tau _2} \left(-1+a \tau _2\right)}{a A \left(-1+4 a \tau _2\right)}e^{a \tau_2}\sim\frac{3 W_0 \sqrt{\tau _2} }{4 a A}e^{a \tau_2}\left(1-\frac{3}{4 a \tau_2}\right),
\label{eq:LVSMin1}
\ee
\be
\tau_2^{3/2}=\frac{\xi  \left(-1+4 a \tau _2\right){}^2}{16  g_s^{3/2} a \tau _2 \left(-1+a \tau _2\right)}\sim\frac{\xi}{g_s^{3/2}}\left(1+\frac{1}{2 a \tau_2}\right).
\label{eq:LVSMin2}
\ee
This allows us to show the second derivatives of $V_{VLS}$ at the origin of the $b$-plane are given by
\be
\begin{split}
&\frac{\partial^2 V_{LVS}}{\partial b_1^2 }=-\frac{9 \xi ^{1/3} k_{21} W_0^2}{64 \sqrt{g_s} \tau _1^{9/2}}-\frac{3 \xi ^{2/3} k_{21}^2 W_0^2}{32 g_s k_{11} \tau _1^5},\\
&\frac{\partial^2 V_{LVS}}{\partial b_1\partial b_2 }=0,\\
&\frac{\partial^2 V_{LVS}}{\partial b_2^2 }=-\frac{9 \xi ^{1/3} k_{22} W_0^2}{64 \sqrt{g_s} \tau _1^{9/2}}-\frac{3 \xi ^{2/3} k_{22}^2 W_0^2}{32 g_s k_{12} \tau _1^5}.
\end{split}
\ee

We can see that the more natural scenario involves having all $k<0$, such that $\frac{\partial^2 V_{LVS}}{\partial b_i\partial b_j }\ge 0$. In such a case, ghosts are avoided if the magnitudes of the intersection numbers obey 
\be
|k_{11}|>|k_{21}| \sqrt{\tau_2/\tau_1}\qquad,\qquad |k_{12}|>|k_{22}| \sqrt{\tau_2/\tau_1},
\ee
as follows from Eq. (\ref{eq:ghosts}). If these conditions are met the $b$ moduli are consistently stabilised by  a quadratic potential at the origin, considerably simplifying the analysis of the inflationary potential.

\subsection{D5 generated potential\label{sec:D5}}
The full scalar potential following from  Eqs. (\ref{eq:W}) and (\ref{eq:KLVS}) has the following structure
\be
V=V_{LVS}+ V_{1}+ V_{2},
\ee
where the D5 generated contributions take the form
\be
V_1=\frac{3 P  W_0 \xi}{64 a \sqrt{g_s} \tau _1^{9/2} \tau _2}\cos\left[c_1 p_1+c_2 p_2\right]+\frac{3 Q   W_0 \xi}{64 a \sqrt{g_s} \tau _1^{9/2} \tau _2}\cos\left[c_1 q_1+c_2 q_2\right] 
\label{eq:V1}
\ee
and
\be\begin{split}
V_2=&-\frac{P Q }{\tau _1^2}\frac{\left(k_{12} p_1 q_1+k_{11} p_2 q_2\right)}{6 k_{11} k_{12} }\cos\left[c_1 p_1+c_2 p_2-c_1 q_1-c_2 q_2\right]\\
& -\frac{P^2}{\tau _1^2}\frac{ \left(k_{12} p_1^2+k_{11} p_2^2\right)}{12 k_{11} k_{12}}-\frac{Q^2}{\tau _1^2}\frac{ \left(k_{12} q_1^2+k_{11} q_2^2\right)}{12 k_{11} k_{12} }.
\label{eq:V2}\end{split}\ee
 From Eqs. (\ref{eq:VLVS}), (\ref{eq:V1}) and (\ref{eq:V2}) one can read off the volume scaling of each component of V close to the LVS minimum:
\be
V_{LVS}\sim \mathcal{O}\left(- \frac{W_0^2}{\vo^3 \log\vo}\right),\qquad V_1\sim \mathcal{O}\left(\frac{QW_0 +P W_0}{\vo^3 \log\vo}\right),\qquad V_2\sim \mathcal{O}\left(\frac{P Q+P^2 +Q^2}{\vo^{4/3}}\right).
\ee

A feature of LVS moduli stabilisation is that the resulting minimum is non-supersymmetric and AdS. Upon uplifting, the barrier separating the LVS vacuum from decompactification has a height roughly equal to the depth of the original AdS minimum: $V_{barrier}\sim \mathcal{O}\left(- \frac{W_0^2}{\vo^3 \log\vo}\right).$ Since any sort of inflationary energy density will contribute to the F-term potential as an aditional uplift term (see \cite{Conlon:2008cj} for a discussion), classical stability of the LVS vacuum requires
\be
V_{barrier}\sim -V_{LVS}\gg V_{inf}=V_1+V_2.
\ee
This can be seen as a constraint on the magnitude of the D5 brane generated $P$ and $Q$ parameters:
\be
P, Q \ll \frac{W_0}{\vo^{5/6}\sqrt{\log \vo}}.
\label{eq:PQsmall}
\ee
With this hierarchy not only do we ensure the classical stability of the vacuum but we also make it possible to decouple the light $c$-axions from their heavier $b$ partners and from the orientifold even moduli.

The inflationary part of the potential, $V_{inf}=V_1+V_2$, can be written in the form of Eq. (\ref{eq:3CosV}) with the identifications
\be
\Lambda_1^4\equiv\frac{3 P  W_0 \xi}{64 a \sqrt{g_s} \tau _1^{9/2} \tau _2},\qquad \Lambda_2^4\equiv \frac{3 Q  W_0 \xi}{64 a \sqrt{g_s} \tau _1^{9/2} \tau _2},\qquad\Lambda_3^4\equiv-\frac{P Q }{\tau _1^2}\frac{\left(k_{12} p_1 q_1+k_{11} p_2 q_2\right)}{6 k_{11} k_{12} }
\ee
and 
\be
\tilde{f}_1\equiv \left(-\frac{3 g_s k_{11}}{2 \tau _1}+\frac{3 g_s k_{21} \sqrt{\tau _2}}{2 \tau _1^{3/2}}\right)^{1/2}\qquad\text{,}\qquad \tilde{f}_2\equiv\left(-\frac{3 g_s k_{12}}{2 \tau _1}+\frac{3 g_s k_{22} \sqrt{\tau _2}}{2 \tau _1^{3/2}}\right)^{1/2}
.
\ee

Due to the fact that the b moduli are stabilised at the origin by $V_{LVS}$, which is independent of whether we want to inflate via  the KNP or the HA mechanisms, the above analysis applies to both cases.

In tables \ref{tab:KNP_in} and \ref{tab:HA_in} we present the input parameters for the KNP and HA numerical examples, whose compactification and inflationary parameters are displayed in table \ref{tab:KNPHA_out}. Semi-realistic models require two distinct tunings, as mentioned above. One starts by choosing the superpotential exponents and the intersection numbers such that the largest effective decay constant is $\ge 5 M_P$. Then one chooses the stabilised volume  and the $P$ and $Q$  parameters to set the scales in the inflationary potential. While doing this one must ensure stability of the LVS vacuum, which requires tuning $P$ and $Q$ small as in Eq. (\ref{eq:PQsmall}). In these models tuning the gauge groups ranks large and  $P$ and $Q$ small seems unavoidable.

\begin{table}[h]
\begin{center}
\begin{tabular}{c||c|c|c|c|c|c|c|c|c|c|c|c|c|c|c}
& $W_0$ & $A$ & $a$ & $P$ & $Q$ & $p_1$&  $q_1$ & $r$ & $\delta$ & $k_{11}$ &$k_{12}$ &$k_{21}$ & $k_{22}$ & $g_s$ & $\xi$\\
\hline
\hline
KNP$_1$ &$1 $&$1$&$\pi$&$10^{-4}$&$10^{-3}$&$\frac{\pi}{2}$&$\frac{\pi}{2}$&$1$&$0.01$&$-3$&$-3$&$-3$&$-3$&$0.3$ &$0.5$\\
\hline
KNP$_2$ &$8$&$2$&$\frac{2\pi}{5}$&$0.01$&$0.01$&$\frac{\pi}{5}$&$\frac{\pi}{5}$&$0.8$&$0.1$&$-6$&$-6$&$-6$&$-6$&$0.3$&$1$\\
\end{tabular}
\end{center}
\caption{Input parameters for the LVS string embedding.}
\label{tab:KNP_in}
\end{table}
\begin{table}[h]
\begin{center}
\begin{tabular}{c||c|c|c|c|c|c|c|c|c|c|c|c|c|c|c}
& $W_0$ & $A$ & $a$ & $P$ & $Q$ & $p_1$&  $p_2$ & $q_1$ & $q_2$ & $k_{11}$ &$k_{12}$ &$k_{21}$ & $k_{22}$ & $g_s$ & $\xi$\\
\hline
\hline

HA$_1$ &$ 1$&$1$&$\frac{2\pi}{5}$&$10^{-3}$&$10^{-3}$&$0$&$\frac{\pi}{11}$&$\frac{\pi}{11}$&$\pi$&$-1$&$-1$&$-1$&$-1$&$0.4$ &$1.5$\\
\hline
HA$_2$ &$10$&$2$&$\pi$&$5\times 10^{-3}$&$5\times 10^{-3}$&$0$&$\frac{2\pi}{15}$&$\frac{2\pi}{25}$&$\pi$&$-10$&$-10$&$-1$&$-1$&$0.5$&$0.7$\\
\end{tabular}
\end{center}
\caption{Input parameters for the LVS string embedding.}
\label{tab:HA_in}
\end{table}

\begin{table}[h]\
\begin{center}
\begin{tabular}{c||c|c||c|c|c|c}
& $\tau_1$ & $\tau_2$ &  $\Lambda_1^4\times 10^{12}$&$\Lambda_2^4\times 10^{12}$&$\Lambda_3^4\times 10^{12}$ & $f_{eff}$\\
\hline
\hline
KNP$_1$ &$48.2$&$2.22$&$0.016$&$0.16$&$11.8$&$13.4$\\
\hline
KNP$_2$ &$51.3$&$3.64$&$30$&$30$&$709$&$5.0$\\
\hline
HA$_1$ &$19.3$&$3.59$&$41$&$41$&$443$&$5.1$\\
\hline
HA$_2$ &$42.3$&$1.70$&$21$&$21$&$306$&$11.23$\\
\end{tabular}
\end{center}
\caption{Compactification features and inflationary parameters. Dimensionful quantities in units of $M_P$.}
\label{tab:KNPHA_out}
\end{table}

\subsection{ED3/ED1 generated potential}

An interesting modification to the scenario analysed above is to consider that the inflationary potential is generated by ED3/ED1 terms in the superpotential rather then by gaugino condensation on D5 branes. In this case the superpotential is 
\be
W=W_0+A e^{-a T_2} \left(1+ P\ e^{- p_1 G_1 - p_2 G_2}+Q\ e^{- q_1 G_1 - q_2 G_2}\right).
\label{eq:WED3ED1}
\ee
The benefit of this setup is that the less volume suppressed part of the  inflationary potential is now down by an extra power of the compactification volume, which alleviates the tuning in the $P$ and $Q$ coefficients. This can be seen by comparing Eqs. (\ref{eq:W}) and (\ref{eq:WED3ED1}) noting that the odd axion part of $W$ receives an extra $e^{-a T_2}\sim \frac{1}{\vo}$ suppression.

In the limit in which one can decouple the c axions from the remaining heavy fields, the minimisation of the orientifold even and b moduli caries over from the previous section, with the vacuum defined by $b_1=b_2=0$ and by Eqs. (\ref{eq:LVSMin1}) and (\ref{eq:LVSMin2}). The dominant contributions to the $c$ axion potential now originate from terms of the form $\left(\frac{\partial W }{\partial T_2}\right)^2 + W_0 \frac{\partial W}{\partial T_2}$, whereas in the D5 case they came from $\left(\frac{\partial W }{\partial G_i}\right)^2 + W_0 \frac{\partial W}{\partial G_i}$. This is the reason why taking Eq. (\ref{eq:WED3ED1}) instead of  (\ref{eq:W}) does not simply imply an extra overall $1/\vo$ suppression. Explicit computation of the potential yields $V=V_{LVS}+V_1+V_2$ with
\be
V_1=-\frac{9 P \xi   W_0^2}{64 a \sqrt{g_s} \tau _1^{9/2} \tau _2}\cos\left[c_1 p_1+c_2 p_2\right]-\frac{9 Q \xi W_0^2}{64 a \sqrt{g_s} \tau _1^{9/2} \tau _2} \cos\left[c_1 q_1+c_2 q_2\right] 
\label{eq:V1ED3ED1}
\ee
and
\be
V_2=\frac{3 \left(P^2+Q^2\right) \xi  W_0^2}{32 \sqrt{g_s} \tau _1^{9/2}}+\frac{3 P Q \xi  W_0^2}{16 \sqrt{g_s} \tau _1^{9/2}}\cos\left[c_1 p_1+c_2 p_2-c_1 q_1-c_2 q_2\right], 
\label{eq:V2ED3ED1}
\ee
from which we can read the leading volume scaling
\be
V_{LVS}\sim \mathcal{O}\left(- \frac{W_0^2}{\vo^3 \log\vo}\right),\qquad V_1\sim \mathcal{O}\left(\frac{QW_0^2 +P W_0^2}{\vo^3 \log\vo}\right),\qquad V_2\sim \mathcal{O}\left(\frac{W_0^2(P Q+P^2 +Q^2)}{\vo^{3}}\right).
\ee
As in the D5 case, vacuum stability requires $V_1, V_2 \ll V_{LVS}$, which now implies 
\be
P, Q \ll \frac{1}{\sqrt{\log\vo}},
\ee
constituting a less severe bound than that of Eq. (\ref{eq:PQsmall}).

The tuning required to obtain trans-Planckian decay constants is exactly as before, apart from the different volume scaling, Eqs. (\ref{eq:V1ED3ED1}) and (\ref{eq:V2ED3ED1}) are identical to (\ref{eq:V1}) and (\ref{eq:V2}) respectively. 

To illustrate the advantages of this setup consider the first numerical example of the KNP mechanism presented in the previous section: KNP$_1$ in table \ref{tab:KNP_in}. To ensure vacuum stability we have required that $|V_{LVS}|\ge 10\ |V_1+V_2|$, this in turn required $(P,Q)=(10^{-4},10^{-3})$. Taking the same criterium for vacuum stability, with the ED3/ED1 generated potential one gets instead $(P,Q)=(10^{-2},3\times 10^{-2})$, a considerable decrease in the level of fine tuning in the model. A similar exercise reveals that one gains a factor of $\mathcal{O}(10-100)$ in the tuning of $P$ and $Q$ in the remaining examples of section \ref{sec:D5} \footnote{Here we are comparing different realisations of inflation at constant volume. By keeping $\vo$ constant and changing $P$ and $Q$ we are increasing the scales of $V_1$ and $V_2$. Realistically these scales are to be set by the COBE normalisation of curvature perturbations once we fix the large effective decay constant. We therefore expect that everything else being equal, by taking ED3/ED1 one can get inflation at the right scale for compactifications with larger $P$ and $Q$ as well as larger $\vo$.}.

\section{Discussion}
Reliable string theoretic large field models of inflation have been sought for a long time, especially if the BICEP2 result holds.
Natural inflation is an elegant model of inflation due to the shift symmetry which protects its potential from dangerous corrections. It also exhibits an interesting range of predictions for the spectral index $n_s$ and the tensor to scalar ratio $r$ which fits current data. The main challenge of the model, is therefore a plausible embedding in a fundamental theory which will explain the decay constant $f\geq 5\ M_p$, necessary to match observations. In string theory one expects many axions, making natural inflation seemingly natural! Nevertheless, there are two notable obstacles. First, as we explained in the introduction, in string theory the decay constants tend to be parametrically smaller than $M_p$. Second, these axions are intricately coupled to moduli fields. Hence, as we explained one must stabilize all moduli and make sure that they do not spoil slow-roll inflation, and that inflation does not destabilize them. Previous works \cite{Kim:2004rp,Ben-Dayan:2014zsa,Tye:2014tja} demonstrated that two fundamental axions produce such an effective decay constant on the field theory level, clearing the first obstacle. This work, passes the second hurdle.

We gave a full account on how to embed the KNP and HA suggestions in a string theory derived setting, that takes into account moduli stabilization.
We have used the $C_2$ axions as our would be inflatons. The inflationary observables follow the pattern of the original natural inflation model. We have demonstrated that the hierarchy needed for moduli stabilization is the same hierarchy needed for the hierarchical axions scenario. Therefore, the KNP and HA scenario, are readily combined with moduli stabilization, in a rather economical way. 

Let us reiterate the main advantages of the models: First inflation is generated only non-perturbative effects, protecting it from perturbative corrections. Second, the smallest
number of axions. Third, there is no manifest tuning of
the input parameters, since only simple hierarchies are necessary. Fourth, the inflationary trajectory is contained in a very small domain $\ll M_p$, again protecting it from dangerous corrections. This sheds light on the large field vs. small field discussion, since it shows that the concern for large corrections due to large field excursions is important only if we consider the inflaton as a fundamental field, while for low energy effective fields, large field excursions are trivial and can be embedded in the constrained string theory setting. Last, it seems that the mathematical structure of the KNP/HA models has a broader context and it will be interesting to investigate this structure thoroughly.

\section*{Acknowledgements}
This work is supported by the Impuls und Vernetzungsfond of the Helmholtz Association of German Research Centres under grant HZ-NG-603, and the German Science Foundation (DFG) within the Collaborative Research Center 676 ``Particles, Strings and the Early Universe''.

\bibliographystyle{JHEP.bst}
\bibliography{2Axions_v4}

\end{document}